%% file: photons_pp.tex
\documentclass[12pt,a4paper]{article}

\usepackage[utf8]{inputenc}

\usepackage[top=15mm,bottom=20mm,left=15mm,right=15mm]{geometry}

\usepackage[affil-it]{authblk}
\usepackage{amsmath,amssymb}
\usepackage{physics}
\usepackage{graphicx}
\usepackage{hyperref}
\usepackage{setspace}
\usepackage{diagbox}
\usepackage{multirow}
\usepackage{nicematrix}

\usepackage[backend=biber,style=ieee]{biblatex}
\newcommand{\ch}{\chi}

\begin{document}

\title{On production of heavy charged particles\\
  in $\gamma\gamma$ fusion at planned $pp$ colliders}

\author[1]{S.~I.~Godunov}
\author[1]{E.~K.~Karkaryan}
\author[1]{V.~A.~Novikov}
\author[2]{A.~N.~Rozanov}
\author[1]{M.~I.~Vysotsky}
\author[1]{E.~V.~Zhemchugov}

\affil[1]{
  \small I.~E.~Tamm Department of Theoretical Physics, Lebedev Physical
  Institute, \newline
  53 Leninskiy Prospekt, Moscow, 119991, Russia
}
 \affil[2]{
 \small Centre de Physique de Particules de Marseille (CPPM), Aix-Marseille
Universite, CNRS/IN2P3, \newline
 163 avenue de Luminy, case 902, Marseille, 13288, France}

\date{}

\maketitle

\begin{abstract}
  Production of heavy fermions in ultraperipheral collisions
  ($pp\to p+\gamma\gamma+p\to p+\ch^{+}\ch^{-}+p$) and the
  semiexclusive reaction
  ($pp \to p+\gamma\gamma^{*}+X \to p+\ch^{+}\ch^{-}+X$) is
  considered. Differential and total cross sections for the energies
  of the planned $pp$ colliders are presented.
\end{abstract}

\section{Introduction}
\label{sec:introduction}

The Standard Model (SM) of elementary particles perfectly describes
relevant experimental data. Nevertheless, it is certain that it should
be expanded in order to solve problems inherent to the SM (too many
free parameters, the hierarchy problem, the origin of neutrino masses,
strong CP violation) as well as to explain the nature of Dark Matter
and Dark Energy and to construct predictive quantum gravity to say the
least. In the course of the SM expansion new heavy particles are
usually introduced.

If these particles are electrically charged they should be produced in
the $\gamma\gamma$-fusion, $\gamma\gamma^{(*)}\to\ch^{+}\ch^{-}$, and
the cross section of this reaction is determined by the values of the
electric charge and the mass of $\ch$. One popular example of
$\ch^{\pm}$ is chargino --- a mixture of superpartners of charged
Higgs and $W^{\pm}$ bosons.\footnote{In the recent
  paper~\cite{2309.16823}, the CMS collaboration presented the results
  of a search for a long-lived chargino in the LHC 2016-2018
  data. In~\cite{2310.08171}, exclusion intervals of chargino masses
  in some particular models are presented by the ATLAS collaboration.}
In what follows we consider protons accelerated at high energy
$pp$-colliders as a source of colliding photons.

We consider two reactions: ultraperipheral collisions when both
protons remain intact and can be used for event tagging with the help
of forward spectrometers, and the semiexclusive process when only one
proton survives while the second disintegrates. We calculate their
cross sections for the planned colliders: HE-LHC (collision energy 27
TeV), SPPC (70 TeV) and FCC (100 TeV) and compare them with what is
obtainable at the LHC (13 TeV).

In Section~\ref{sec:analytical} necessary formulas are
presented. Numerical results are given in
Section~\ref{sec:numerical}. In Section~\ref{sec:conclusions} we
conclude.

\section{Formulas for the cross sections}
\label{sec:analytical}

One of the necessary ingredients of the calculation is the cross
section of the $\gamma\gamma^{(*)}\to\ch^{+}\ch^{-}$ reaction. Due to
the elastic form factor, virtuality of the photon emitted by the
survived proton $Q_{1}^{2}\equiv -q_{1}^{2}$, where $q_{1}$ is the
photon 4-momentum, is bounded by approximately
$\left(200~\text{MeV}\right)^{2}$~\cite[Appendix A]{1806.07238}. Since
the contribution of longitudially polarized photon is proportional to
$Q_{1}^{2}/W^{2}$ where $W$ is the invariant mass of the produced
pair, it can be safely neglected and only the transverse polarization
of this photon should be taken into account. However, in the case of
the disintegrating proton both transversal and longitudial
polarizations should be accounted for. Formulas for the cross section
of the massive fermions pair production in the collision of a real and
a virtual photons are presented in~\cite[Appendix E,
Eq.~(E3)]{Budnev:1975poe}:
\begin{align}
  \sigma_{TS}&=\frac{4\pi\alpha^{2}}{W^{2}}
               \frac{1}{\left(1+\frac{Q_{2}^{2}}{W^{2}}\right)^{3}}
               \frac{4Q_{2}^{2}}{W^{2}}\left(\sqrt{1-\frac{4m_{\ch}^{2}}{W^{2}}}-\frac{2m_{\ch}^{2}}{W^{2}}
               \ln\frac{1+\sqrt{1-\frac{4m_{\ch}^{2}}{W^{2}}}}{1-\sqrt{1-\frac{4m_{\ch}^{2}}{W^{2}}}}\right),\\
  \sigma_{TT}&=\frac{4\pi\alpha^{2}}{W^{2}}
               \frac{1}{\left(1+\frac{Q_{2}^{2}}{W^{2}}\right)^{3}}
               \left(\left(1+\frac{Q_{2}^{4}}{W^{4}}+\frac{4m_{\ch}^{2}}{W^{2}}-\frac{8m_{\ch}^{4}}{W^{4}}\right)
               \ln\frac{1+\sqrt{1-\frac{4m_{\ch}^{2}}{W^{2}}}}{1-\sqrt{1-\frac{4m_{\ch}^{2}}{W^{2}}}}\right. \\
  \nonumber & \hspace{7cm} \left. -
              \left(\left(1-\frac{Q_{2}^{2}}{W^{2}}\right)^{2}+\frac{4m_{\ch}^{2}}{W^{2}}\right)
              \sqrt{1-\frac{4m_{\ch}^{2}}{W^{2}}}
              \vphantom{\ln\frac{1+\sqrt{1-\frac{4m_{\ch}^{2}}{W^{2}}}}{1-\sqrt{1-\frac{4m_{\ch}^{2}}{W^{2}}}}}\right),
\end{align}
where $\sigma_{TS}$ is the cross section when the photon emitted by
the disintegrating proton is polarized longitudinally, $\sigma_{TT}$
is that when it is polarized transversally, $\alpha$ is the fine
structure constant, $m_{\ch}$ is the mass of $\ch^{\pm}$, $W$ is the
invariant mass of the produced pair, $Q_{2}^{2}$ is the virtuality of
the photon emitted by the disintegrating proton.

In what follows we need the total cross section:
\begin{align}
  \label{eq:cs_full}
  &\sigma_{\gamma\gamma^{*}\to
  \ch^{+}\ch^{-}}\left(W^{2},Q_{2}^{2},m_{\ch}^{2}\right)
  {}\equiv{}
  \sigma_{TS}+\sigma_{TT}=\\ \nonumber
  &\hspace{2cm}=\frac{4\pi\alpha^{2}}{W^{2}}
    \frac{1}{\left(1+\frac{Q_{2}^{2}}{W^{2}}\right)^{3}}
    \left(\left(1+\frac{Q_{2}^{4}}{W^{4}}+\frac{4m_{\ch}^{2}}{W^{2}}-\frac{8m_{\ch}^{4}}{W^{4}}-\frac{8m_{\ch}^{2}Q_{2}^{2}}{W^{4}}\right)
               \ln\frac{1+\sqrt{1-\frac{4m_{\ch}^{2}}{W^{2}}}}{1-\sqrt{1-\frac{4m_{\ch}^{2}}{W^{2}}}}\right. \\
  \nonumber & \hspace{8cm} \left. -
              \left(1-\frac{6Q_{2}^{2}}{W^{2}}+\frac{Q_{2}^{4}}{W^{4}}+\frac{4m_{\ch}^{2}}{W^{2}}\right)
              \sqrt{1-\frac{4m_{\ch}^{2}}{W^{2}}}
              \vphantom{\ln\frac{1+\sqrt{1-\frac{4m_{\ch}^{2}}{W^{2}}}}{1-\sqrt{1-\frac{4m_{\ch}^{2}}{W^{2}}}}}\right).
\end{align}

The cross section of $\ch^{+}\ch^{-}$ pair production in the fusion of
photons emitted by a disintegrating and an elastically scattered
protons equals~\cite[Eqs.~(18)--(21), see also Eq.~(41)]{2308.01169}
\begin{align}
  \label{eq:main}
  \frac{\dd\sigma_{pp\rightarrow p\ch^+\ch^-X}}{\dd W} =
  & \frac{4\alpha W}{\pi} \sum_q Q^2_q
    \int\limits^{s-W^{2}}_{\frac{W^4}{36\gamma^2s}}
    \frac{\sigma_{\gamma\gamma^* \to \ch^+\ch^-}\qty(W^2,
    Q_2^2,m_{\ch}^{2})}{\qty(W^2+Q^2_2)Q^4_2}\cdot
    \dd Q^2_2 \times \\ \nonumber
  & \times
    \int\limits^1_{\frac{W^{2}+Q_{2}^{2}}{s}\cdot\max\qty(1,\frac{m_{p}}{3\sqrt{Q_{2}^{2}}})}\hspace{-2em}\dd x f_q(x, Q^2_2)
    \int\limits^{\frac{1}{2}\ln{\frac{s}{W^2+Q^2_2}}}_{\frac{1}{2}\ln{\qty(\frac{W^2+Q^2_2}{x^2s}\cdot\max\qty(1,\frac{m_{p}^{2}}{9Q_{2}^{2}}))}}
    \hspace{-2em}\omega_1 n_p(\omega_1) \qty[Q^2_2-\qty(\omega_2/3x\gamma)^2] \dd y,
\end{align}
where $q$ is a quark inside the disintegrating proton, $Q_{q}$ is its
charge, $s$ is the square of the invariant mass of the colliding
protons, $m_{p}$ is the proton mass,
$\gamma=\sqrt{s}/\left(2m_{p}\right)$ is the proton Lorenz factor,
$f_{q}\left(x,Q_{2}^{2}\right)$ is the parton distribution function
(PDF) for quark $q$, $n_{p}\left(\omega\right)$ is the equivalent
photon spectrum of proton~\cite[Eqs.~(4) and (5)]{2207.07157},
$y=\left(1/2\right)\ln\left(\omega_{1}/\omega_{2}\right)$ is the pair
rapidity, $\omega_{1}$ and $\omega_{2}$ are photons energies,
$\omega_1 = \sqrt{W^2+Q^2_2}\cdot e^y/2$,
$\omega_2 = \sqrt{W^2+Q^2_2}\cdot e^{-y}/2$.

The accuracy of \eqref{eq:main} is at the level of 20~\% which is
sufficient for our estimates. More details can be found
in~\cite{2308.01169} including the discussion of uncertainties due to
PDFs and the impact of low-$Q^{2}$ physics. The results of this paper
were obtained with the help of \texttt{MMHT2014nnlo68cl} PDF
set~\cite{1412.3989} provided by LHAPDF~\cite{1412.7420}.

For the quasielastic process $pp\to p\ch^{+}\ch^{-}p$ we have (see
Eqs.~(2.15) and (2.16) in~\cite{2106.14842})
\begin{equation}
  \label{eq:quasielastic}
  \frac{\dd\sigma_{pp\rightarrow p\ch^+\ch^-p}}{\dd W} =
  \sigma_{\gamma\gamma^* \to \ch^+\ch^-}\qty(W^2,
    Q_2^2=0,m_{\ch}^{2})\cdot
    \frac{W}{2}\int\limits_{-\infty}^{\infty}
    n_{p}\left(\frac{W}{2}e^{y}\right)
    n_{p}\left(\frac{W}{2}e^{-y}\right) \dd y.
\end{equation}

\section{Pair production at future $pp$ colliders}
\label{sec:numerical}

In this section we consider pair production of the charged fermions
$\ch^{+}\ch^{-}$ in the photon fusion in semiexclusive reactions and
ultraperipheral collisions at the planned $pp$ colliders:
\begin{itemize}
\item HE-LHC (energy $\sqrt{s}=27$~TeV, luminosity $\mathcal{L}=16\cdot 10^{34}~\text{cm}^{-2}\cdot\text{s}^{-1}$)~\cite{FCC:2018bvk};
\item SPPC (energy $\sqrt{s}=70$~TeV, luminosity $\mathcal{L}=12\cdot 10^{34}~\text{cm}^{-2}\cdot\text{s}^{-1}$)~\cite{1507.03224};
\item FCC-hh (energy $\sqrt{s}=100$~TeV, luminosity $\mathcal{L}=5\cdot 10^{34}~\text{cm}^{-2}\cdot\text{s}^{-1}$)~\cite{FCC:2018vvp}.
\end{itemize}
To get an integrated luminosity in one year of operation, one should
multiply these luminosities by~$10^{7}$~s (the following results are
obtained assuming this duration of collecting data at peak
luminosity). For comparison we present results for the LHC with
$\sqrt{s}=13$~TeV and $140~\text{fb}^{-1}$ of integrated luminosity
collected by the ATLAS and the CMS collaborations in the years
2016--2018 ($140~\text{fb}^{-1}=1.4\cdot10^{41}~\text{cm}^{-2}$).

\begin{figure}[p]
  \centering
  \includegraphics[width=7in]{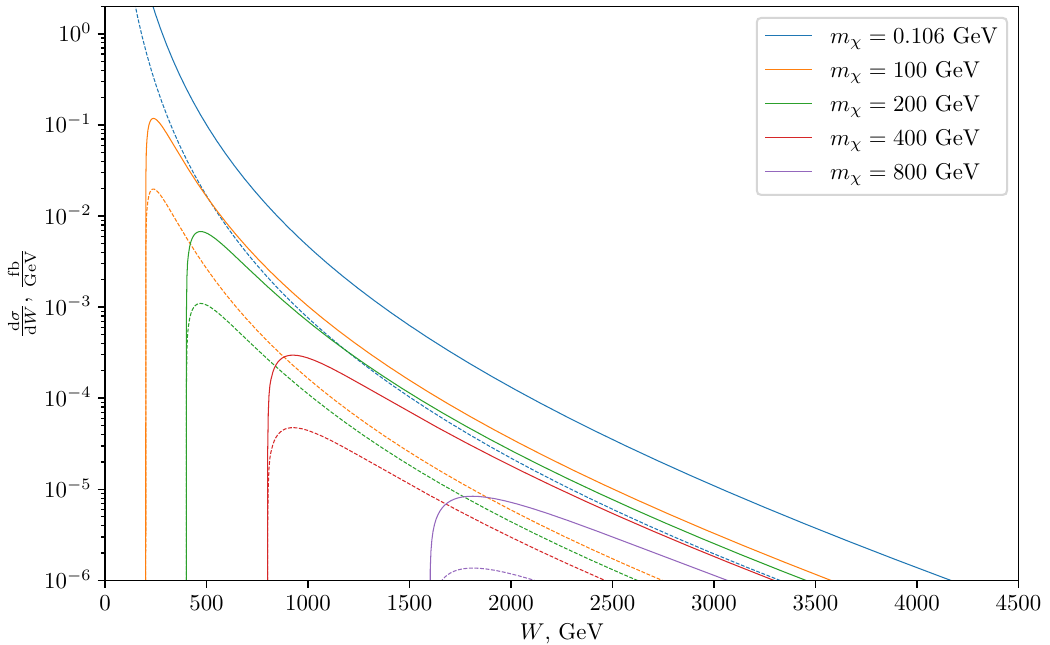}
  \caption{Differential cross sections~\eqref{eq:main} (solid lines)
    and~\eqref{eq:quasielastic} (dashed lines) for
    $\sqrt{s}=13~\text{TeV}$ and $m_{\ch}=0.106$, 100, 200, 400,
    800~GeV (top to bottom).}
  \label{fig:dsigma_13}
  \vspace{5mm}
  \includegraphics[width=7in]{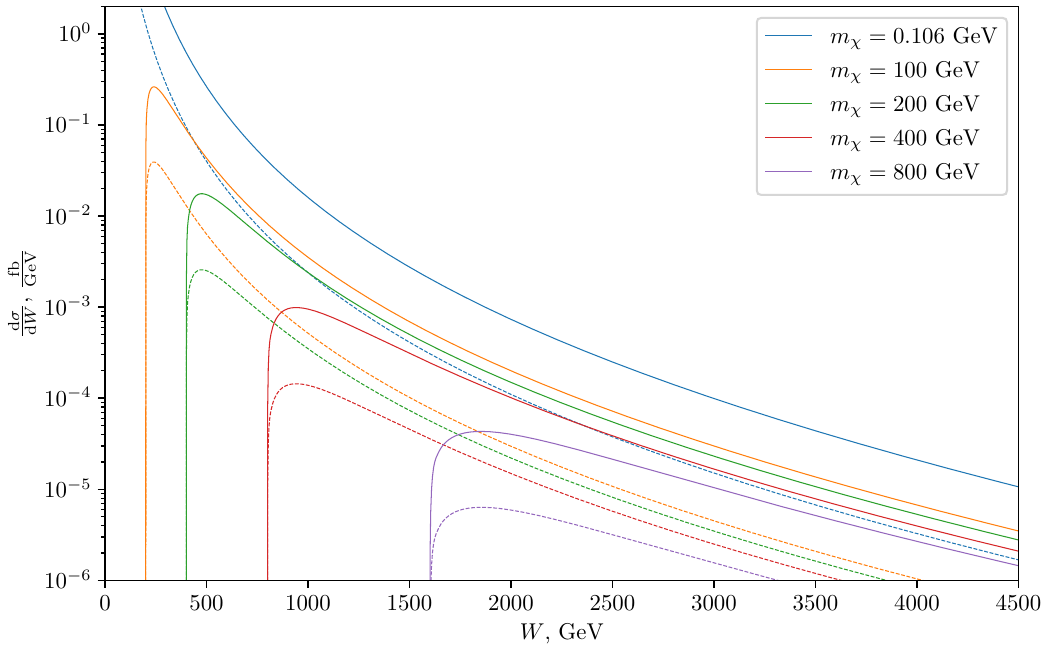}
  \caption{As in Fig.~\ref{fig:dsigma_13}, but for $\sqrt{s}=27$~TeV.}
  \label{fig:dsigma_27}
\end{figure}

\begin{figure}[p]
  \centering
  \includegraphics[width=7in]{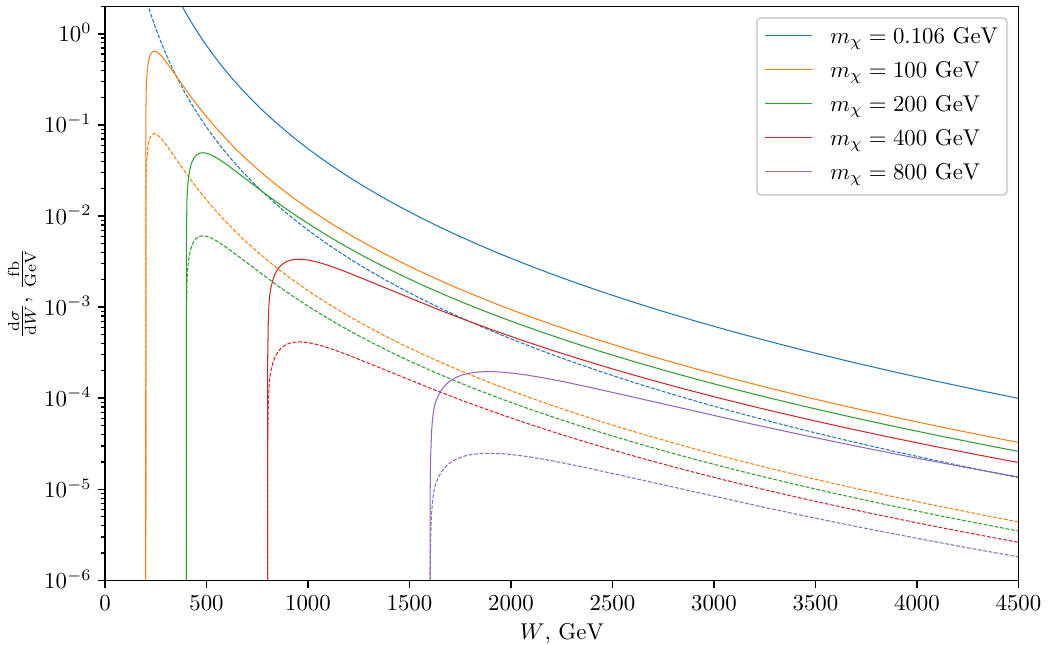}
  \caption{As in Fig.~\ref{fig:dsigma_13}, but for $\sqrt{s}=70$~TeV.}
  \label{fig:dsigma_70}
  \vspace{10.4mm}
  \includegraphics[width=7in]{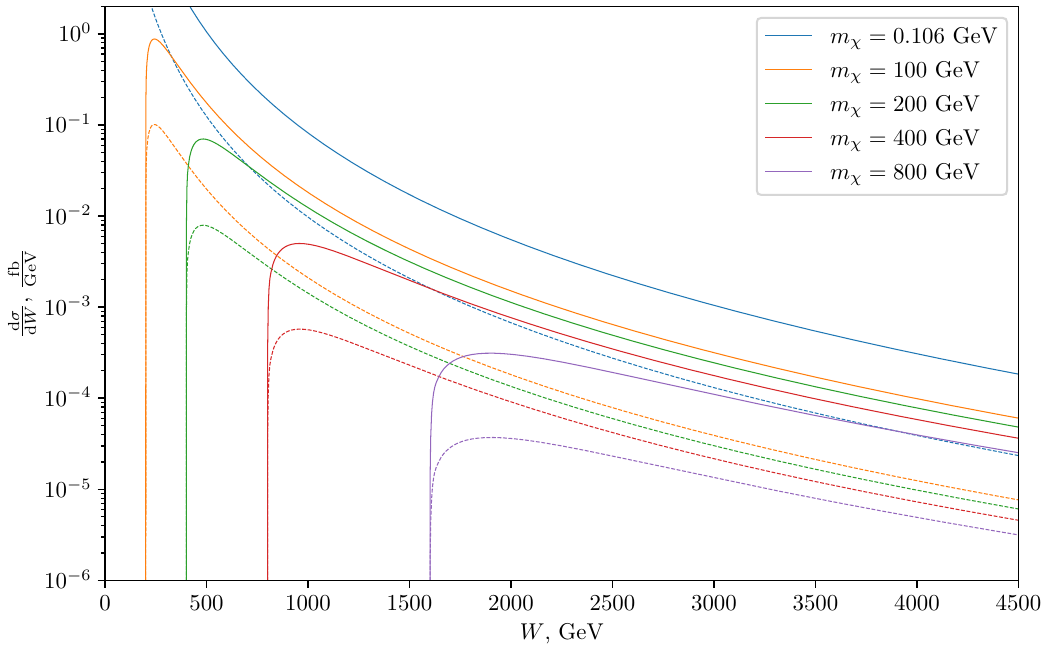}
  \caption{As in Fig.~\ref{fig:dsigma_13}, but for $\sqrt{s}=100$~TeV.}
  \label{fig:dsigma_100}
\end{figure}

Differential cross sections $\dd\sigma/\dd W$ for $\sqrt{s}=13$, 27,
70, 100~TeV are presented in
Figures~\ref{fig:dsigma_13}--\ref{fig:dsigma_100}. Each figure
contains plots for $\ch$ masses 100, 200, 400, 800~GeV as well as for
muon pair production.

\begin{table}[t]
  \centering
  \begin{NiceTabular}{|W{c}{12mm}*{10}{|c}|}
    \hline
    \Block{3-1}{
    \diagbox{
      $\sqrt{s}$,\\ TeV
    }{
      $m_{\ch}$,\\ GeV
    }}
    & \Block{2-2}{0.106}&
    & \Block{2-2}{100}&
    & \Block{2-2}{200}&
    & \Block{2-2}{400}&
    & \Block{2-2}{800}& \\
    &&&&&&&&&& \\
    \cline{2-11}
    & UPC & SE
    & UPC & SE
    & UPC & SE
    & UPC & SE
    & UPC & SE \\
    \hline
    \input{full.tex}\\
    \hline
  \end{NiceTabular}
  \caption{Total cross sections (in fb) for $\ch^{+}\ch^{-}$ pair
    production in ultraperipheral collisions
    \mbox{$pp\to p\ch^+\ch^-p$} (UPC; integral
    of~\eqref{eq:quasielastic}) and in the inelastic process
    \mbox{$pp\to p\ch^+\ch^-X$} (SE; integral of~\eqref{eq:main}). The
    column with $m_{\ch}=0.106$~GeV corresponds to muon pair
    production with the threshold $W>12$~GeV.}
  \label{tab:total}
\end{table}

Total cross sections are collected in Table~\ref{tab:total}. In the
case of muons we integrate over $W>12$~GeV since this lower bound was
implemented in~\cite{1708.04053} in order to suppress the
background. Cross sections of muon pair production for
$\sqrt{s}=13$~TeV and $W>12$~GeV were already calculated
in~\cite{2112.01870}: 203~pb and 60~pb for the inelastic and the
quasielastic cross sections correspondingly. The small differences are
due to additional approximations made in~\cite{2112.01870} (the most
notable of these is neglecting the Pauli form factor contribution in
the equivalent photons spectrum of proton).

It was shown in~\cite{1906.08568} that for the quasielastic processes
the background can be almost eliminated with the help of proton
tagging. So the discovery potential is mostly defined by the number of
events. Though the designs of the future experiments are different
(and therefore selection criteria will also be different), we can
compare total numbers of produced pairs assuming similar efficiencies
of event selection. These numbers are presented in
Table~\ref{tab:total_N}.
\begin{table}[t]
  \centering
  \begin{NiceTabular}{|W{c}{12mm}*{10}{|c}|}
    \hline
    \Block{3-1}{
    \diagbox{
      $\sqrt{s}$,\\ TeV
    }{
      $m_{\ch}$,\\ GeV
    }}
    & \Block{2-2}{0.106}&
    & \Block{2-2}{100}&
    & \Block{2-2}{200}&
    & \Block{2-2}{400}&
    & \Block{2-2}{800}& \\
    &&&&&&&&&& \\
    \cline{2-11}
    & UPC & SE
    & UPC & SE
    & UPC & SE
    & UPC & SE
    & UPC & SE \\
    \hline
    \input{full_N.tex}\\
    \hline
  \end{NiceTabular}
  \caption{Total number of events for $\ch^{+}\ch^{-}$ pair production
    in ultraperipheral collisions \mbox{$pp\to p\ch^+\ch^-p$} (UPC;
    integral of~\eqref{eq:quasielastic}) and in the inelastic process
    \mbox{$pp\to p\ch^+\ch^-X$} (SE; integral of~\eqref{eq:main}). The
    column with $m_{\ch}=0.106$~GeV corresponds to muon pair
    production with the threshold $W>12$~GeV. While for the LHC we
    take the available integrated luminosity 140~fb$^{-1}$, for the
    fulture colliders we assume one year of operation ($10^{7}$~s) at
    expected luminosity.}
  \label{tab:total_N}
\end{table}

In~\cite{1906.08568} it was shown that heavy charged fermions with the
mass up to almost 200~GeV can be found at $3\sigma$ level in the
process $pp\rightarrow p\ch^+\ch^-p$ with the help of LHC Run-2
data. We see from Table~\ref{tab:total_N} that we should have a
similar number of events for $m_{\ch}=800$~GeV at the SPPC. Therefore
it is possible to push the model independent lower bound on heavy
charged fermions mass to about 800~GeV, or discover these new
particles. The SPPC has the greatest potential due to larger expected
luminosity.

\section{Conclusions}
\label{sec:conclusions}

Cross sections for pair production of heavy charged particles in both
inelastic $pp\rightarrow p\ch^+\ch^-X$ and quasielastic
$pp\rightarrow p\ch^+\ch^-p$ processes were calculated for the future
$pp$ colliders. Total numbers of events were estimated based on the
expected luminosity of these experiments. The SPPC has the greatest
potential and can find heavy charged fermions with masses up to about
800~GeV in one year of operation. Let us stress that there are many
more semiexclusive events than quasielastic ones.

The main advantage of the considered processes is the possibility to
detect survived proton(s) which provides effective means for
background suppression. Nowadays, when the detectors for these
colliders are intensively discussed, we would like to emphasize the
importance of forward spectrometers that could provide unique model
independent methods for Beyond Standard Model searches.

Numerical results were obtained with the help of \texttt{libepa}
library~\cite{2311.01353}.

We are supported by RSF grant No. 19-12-00123-$\Pi$.

\printbibliography

\end{document}

%% file: full.tex
13 & 69100 & 229000 & 3.45 & 20.9 & 0.341 & 2.11 & 0.0253 & 0.157 & 0.00117 & 0.00697 \\
27 & 102000 & 367000 & 7.46 & 50.4 & 0.901 & 6.17 & 0.0903 & 0.617 & 0.00681 & 0.0458 \\
70 & 158000 & 638000 & 16.6 & 134 & 2.36 & 19.2 & 0.301 & 2.4 & 0.0326 & 0.253 \\
100 & 184000 & 772000 & 21.4 & 187 & 3.19 & 27.9 & 0.433 & 3.7 & 0.0514 & 0.424

%% file: full_N.tex
13 & $9.67\cdot 10^{6}$ & $3.21\cdot 10^{7}$ & 483 & 2920 & 47.8 & 296 & 3.54 & 22.0 & 0.164 & 0.976 \\
27 & $1.63\cdot 10^{8}$ & $5.87\cdot 10^{8}$ & 11900 & 80700 & 1440 & 9870 & 144 & 987 & 10.9 & 73.3 \\
70 & $1.90\cdot 10^{8}$ & $7.66\cdot 10^{8}$ & 19900 & 161000 & 2830 & 23000 & 361 & 2880 & 39.2 & 304 \\
100 & $9.20\cdot 10^{7}$ & $3.86\cdot 10^{8}$ & 10700 & 93700 & 1590 & 14000 & 216 & 1850 & 25.7 & 212